\documentclass[%
 reprint,
nofootinbib,
 amsmath,amssymb,
 aps,
]{revtex4-2}

\usepackage{graphicx}
\usepackage{dcolumn}
\usepackage{bm}
\usepackage{hyperref}
\hypersetup{
    colorlinks=true,
    linkcolor=blue,
    filecolor=magenta,      
    urlcolor=cyan,
    pdfpagemode=FullScreen,
}
\usepackage{color,soul}

\usepackage{soul}

\usepackage{color}

\begin{document}

\preprint{APS/123-QED}

\title{Comprehensive study of compact stars with dark matter}

\author{Mikel F. Barbat}
 \email{mfernandez@ice.csic.es}
 
\affiliation{Institute of Space Sciences (ICE, CSIC), Campus UAB, Carrer de Can Magrans, 08193 Barcelona, Spain}

\author{J\"urgen Schaffner-Bielich}
\email{schaffner@astro.uni-frankfurt.de}
\affiliation{Institut f\"ur Theoretische Physik, J. W. Goethe Universit\"at, 
  Max von Laue-Str.~1, 60438 Frankfurt am Main, Germany}

\author{Laura Tolos}
 \email{tolos@ice.csic.es}
\affiliation{Institute of Space Sciences (ICE, CSIC), Campus UAB, Carrer de Can Magrans, 08193 Barcelona, Spain}
\affiliation{Institut d'Estudis Espacials de Catalunya (IEEC), 08860 Castelldefels (Barcelona), Spain}
\affiliation{Frankfurt Institute for Advanced Studies, Ruth-Moufang-Str. 1, 60438 Frankfurt am Main, Germany}

\date{\today}

\begin{abstract}
We present a comprehensive study of compact stars admixed with non-self annihilating self-interacting fermionic dark matter, delineating the dependence on the nuclear equation of state by considering the two limiting parametrized equations of state for neutron star matter obtained by smoothly matching the low-density chiral effective theory and the high-density perturbative QCD. These two parametrizations are the limiting cases of a wide variety of smooth equations of state, i.e.\ the softest and stiffest possible one without a phase transition, that generate masses and radii compatible with 2M$_\odot$ observations and the tidal constraint from GW170817. With an exhaustive analysis of the possible stable mass-radius configurations, we determine the quantity of dark matter contained in stars with masses and radii compatible with the aforementioned astrophysical constraints. We find that for dark particle masses of a few tenths of GeV, the dark core collapses and no stable solutions are found  for the two limiting ordinary matter equations of state. For lower masses, the dark matter fraction is limited to 10\%, being at most 1\% for masses ranging from 0.1 to 10 GeV for the limiting soft nuclear equation of state. For the limiting stiff nuclear equation of state, the dark matter fraction can reach values of more than 10\%, but the dark particle mass is being constrained to 0.3 GeV and 10 GeV for the weak self-interacting case and has to be at least 5 GeV for the strong self-interacting one. For dark particle masses of less than 0.1 GeV, stable neutron star configurations should have less than 1\% of self-interacting dark matter to be compatible with the constraint of the tidal deformability from GW170817  for the two limiting ordinary matter equations of state studied.

\end{abstract}

\maketitle

\section{Introduction}

According to astrophysical and cosmological observations,  \cite{Planck2018,SDSS:2014iwm,KHLOPOV2021103824}, most of the mass present in our Universe would appear as dark matter (DM). DM is a hypothetical form of matter that does not absorb or reflect light, whose detection and nature are still evasive. Several methods have been developed to find signatures of DM, such as the direct methods involving particle accelerators \cite{ATLAS:2012ky,CMS:2012ucb} or direct searches based on the DM scattering off nuclear targets in detectors  \cite{Klasen:2015uma}. Also, it has been postulated the possibility of detecting DM and extracting information on their properties by means of analysing its effect on compact stars, such as neutron stars. 

Neutron stars are one of the most compact objects in our Universe, whose masses can exceed $2M_{\odot}$ \cite{Demorest2010ShapiroStar,antoniadis2013massive,Fonseca2016,cromartie2020relativistic} and with radii around $11-15$ km (see latest results from the NICER collaboration~\cite{Riley:2019yda,Miller:2019cac,Riley:2021pdl,Miller:2021qha}). Among the different indirect searches of DM in neutron stars, the authors of Refs.~\cite{Goldman:1989nd,Kouvaris:2011fi,Fuller:2014rza,Acevedo:2020gro,Ray:2023auh,Bhattacharya:2023stq,Bhattacharya:2024pmp} have studied the gravitational collapse of a neutron star due to DM accretion so as to set bounds on the DM properties. Moreover, in Ref.~\cite{Kouvaris:2010jy} the accretion of DM in Sun-like or supermassive stars and the subsequent collapse into a neutron star or a white dwarf has been analysed. Also,  there have been studies about the modification of the cooling pattern of compact stars due to the presence of DM, which will finally self-annihilate \cite{Kouvaris:2007ay,Bertone:2007ae,Kouvaris:2010vv,McCullough:2010ai,deLavallaz:2010wp,Sedrakian:2018kdm,Bhat:2019tnz}, and works on the changes in the kinematical properties of neutron stars due to the accretion of self-annihilating DM \cite{PerezGarcia:2011hh}. 
 
Neutron stars that accummulate DM have also emerged as interesting scenarios to analyze the effects of DM onto hadronic (or quark) matter \cite{1983SvA....27..371B,1991SvA....35...21K,deLavallaz:2010wp,Li:2012ii,Sandin:2008db,Leung:2011zz,Leung:2012vea,Xiang:2013xwa,Goldman:2013qla,Khlopov:2013ava,Mukhopadhyay:2015xhs,tolos2015dark,Rezaei:2016zje,Panotopoulos:2017idn,Nelson:2018xtr,Ellis:2018bkr,Gresham:2018rqo,Deliyergiyev:2019vti,DelPopolo:2019nng,Ivanytskyi:2019wxd,Dengler:2021qcq, Karkevandi:2021ygv,Sen:2021wev,Guha:2021njn,Karkevandi:2021ygv,Miao:2022rqj,Sen:2022pfr,Ferreira:2022fjo,Shakeri:2022dwg,Hippert:2022snq,Cassing:2022tnn,Zollner:2022dst,Collier:2022cpr,Lenzi:2022ypb,Zollner:2023myk,Cronin:2023xzc,Thakur:2023aqm,Sagun:2023rzp,Mariani:2023wtv,Diedrichs:2023trk,Liu:2023ecz,Bramante:2023djs,Giangrandi:2024qdb,Guha:2024pnn,Karkevandi:2024vov,liu2024dark}. In these works the masses and radii of these compact stellar objects (as well as, in some cases, dynamical processes, such as cooling) have been thoroughly studied assuming fermionic or bosonic DM that interacts gravitationally with ordinary neutron star matter (OM) or, in certain studies, through the weak force. In this context, in Ref.~\cite{tolos2015dark} some of us have proposed the possible existence of compact stars that contain DM with Earth or Jupiter-like masses but unusual radii, the so-called dark compact planets.  

Moreover, since the GW170817 gravitational-wave event of a binary neutron-star system \cite{LIGOScientific:2017vwq,Abbott_2018,Abbott_2019}, the possible signature of DM in mergers have been also addressed, specially in the post-merger phase \cite{Ellis:2017jgp,Bezares:2019jcb,Horowitz:2019aim,Bauswein:2020kor}. As for the inspiral phase, the GW signal could be modified due the changes in the structure of neutron stars induced by the presence of DM. Indeed, several studies have been performed on the modifications of the second Love number, and hence, tidal deformability due to the possible existence of DM~\cite{Nelson:2018xtr,Ellis:2018bkr,Ivanytskyi:2019wxd,Dengler:2021qcq,Sen:2021wev,Collier:2022cpr,Lenzi:2022ypb,Thakur:2023aqm,Mariani:2023wtv,Diedrichs:2023trk,Karkevandi:2024vov,liu2024dark}. In particular, in a previous work Ref.~\cite{Dengler:2021qcq} it was found that the second Love numbers for neutron stars admixed with DM are markedly different compared to those expected for neutron stars without DM.

In the present paper we perform an exhaustive study of the properties of compact stars admixed with non-self annihilating self-interacting DM. To this end, we make use of two different EoS parametrizations for OM \cite{Kurkela:2014vha} that fulfil the well-known limits at low and large nuclear densities while giving rise to masses and radii that span the possible mass-radius region compatible with 2$M_{\odot}$ observations and the GW170817 constraint on the tidal deformability, thus, being the limiting cases of a wide variety of EoS that fall between these two extreme cases. We first perform an extensive study of possible stable mass-radius configurations, obtaining solutions away from the typical values for neutron stars. We then find out the quantity of DM contained in stars with masses and radii compatible with the aforementioned astrophysical constraints in terms of the mass of the DM particle and the strength of DM self-interaction. 

This work goes beyond our previous studies of Refs~\cite{tolos2015dark,Dengler:2021qcq} in two ways. On the one hand, our results are obtained using two EoS parametrizations that represent a wide variety of EoSs, going beyond the use of a specific model. On the other hand, we analyze the stability of mass-radius configurations following the recent procedure described in Ref.~\cite{Hippert:2022snq}, that considers the changes in stability that OM might induce on DM, and viceversa, hence outdoing our previous naive study based on the separate stability analysis of DM and OM. In this manner, our results are independent on the model used for OM and grounded on a formal generalisation of the stability criterion for compact stars made of OM admixed with DM.

 Moreover, we should compare our investigation with the recent studies of Refs.~\cite{Ivanytskyi:2019wxd,Mariani:2023wtv,Karkevandi:2024vov,liu2024dark}. Whether the present and former works aim at determining the accumulated DM fraction compatible with astrophysical observations, some important differences can be drawn among the different works. The conclusions of Refs.~\cite{Ivanytskyi:2019wxd,Karkevandi:2024vov,liu2024dark} are based on a individual model, either fermionic or bosonic DM, whereas the outcome of Ref.~\cite{Mariani:2023wtv} and of our present analysis are grounded on two limiting EoS parametrizations that enclose a wide variety of EoSs. Also, the authors of Refs.~\cite{Ivanytskyi:2019wxd,Mariani:2023wtv} only consider mass-radius configurations close to the neutron-star solutions, and no stability analysis for the different configurations has been performed in none of Refs.~\cite{Ivanytskyi:2019wxd,Mariani:2023wtv,Karkevandi:2024vov,liu2024dark}, thus in stark contrast with our present work where an exhaustive stability analysis is carried out for mass-radius configurations close and beyond typical mass-radius for neutron stars. Note that a detailed comparison among the different studies will be also presented when discussing our results.

The paper is organized as follows. In Sec.~\ref{sec:formalism} we introduce the Tolman-Oppenheimer-Volkov (TOV) for two fluids and the stability criterion for the different mass-radius configuration. In Sec.~\ref{sec:results} we show the results for the masses, radii and tidal deformabilities for the different stable configurations, whereas we determine the quantity of DM that can be accumulated on compact stars depending on the DM mass particle and the DM self-interacting strength. Finally, in Sec.~\ref{sec:summary} we present our conclusions and outlook.

\section{Formalism}
\label{sec:formalism}

\subsection{TOV-equations for two fluids}

In this work we investigate compact objects that
are made of OM admixed with non-self-annihilating self-interacting DM. These types of matter are represented by two fluids that only interact gravitationally. We will follow the work of Ref.~\cite{Sandin:2008db}, where the TOV equations for two fluids are determined. As done in Refs.~\cite{tolos2015dark,Dengler:2021qcq}, we use the modified dimensionless TOV equations to obtain the masses and radii  

\begin{eqnarray}
    &&\dfrac{dp'_{OM}}{dr}=-(p'_{OM}+\epsilon'_{OM}) \dfrac{d \nu}{dr}, \nonumber\\
    &&\dfrac{dm_{OM}}{dr}=4\pi r^2 \epsilon'_{OM}, \nonumber \\
    &&\dfrac{dp'_{DM}}{dr}=-(p'_{DM}+\epsilon'_{DM}) \dfrac{d \nu}{dr},\nonumber \\
    &&\dfrac{dm_{DM}}{dr}=4\pi r^2 \epsilon'_{DM},\nonumber \\
    &&\dfrac{d \nu}{dr} = \dfrac{(m_{OM}+m_{DM})+4\pi r^3 (p'_{OM}+p'_{DM})}{r(r-2(m_{OM}+m_{DM})},
\label{eq:TOV}
\end{eqnarray}
where $p'_i$ and $\epsilon '_i$ are the dimensionless pressure and energy density for each species ($i= \{ {\rm OM}, \rm{DM} \}$), defined as $p'_i=p_i /m_f^4$ and $\epsilon'_i=\epsilon_i /m_f^4$, where we have chosen the fermionic DM particle ($m_f$) as the common rescaling. In this manner we can solve the TOV equations and obtain the physical mass $M_i$ and radius $R_i$ for each species after being re-scaled by $M_i=(M_p^3/m_f^2)m_i$ and $R_i=(M_p/m_f^2)r_i$, where $M_p$ is the Planck mass~\cite{Narain:2006kx}.

As mentioned in the Introduction, the aim of this paper is to further analyse the impact of DM on the mass, radius and tidal deformability of compact objects made of OM and DM, starting from a controlled model for OM.

\begin{figure}[!h]
    \centering
    \includegraphics[scale=0.395]{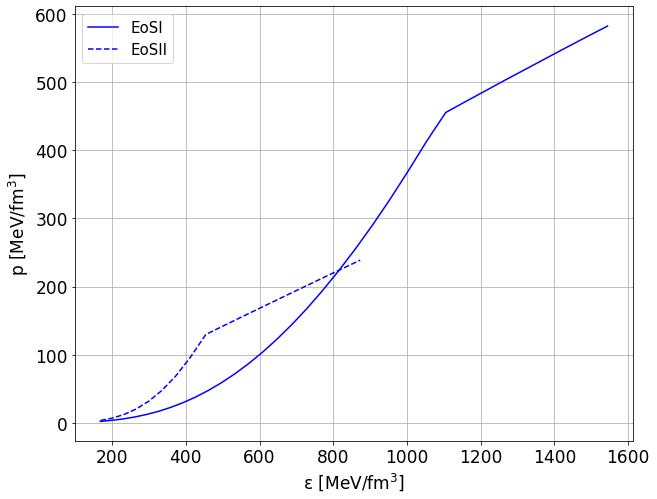}
    \caption{EoSI and EoSII used for the high density matter in the neutron star core \cite{Kurkela:2014vha}.}
    \label{fig:EoSOM}
\end{figure}

 Thus, we consider two different EoSs for OM that result from the causal and smooth polytropic interpolation between the low-density chiral effective theory and the high-density perturbative QCD, and give rise to masses and radii that are at the borderlines for the possible mass-radius region compatible with $2\rm{M}_\odot$ \cite{Kurkela:2014vha}, hence being representative of a wide variety of smooth EoSs that fall between these two limiting cases. In Ref.~\cite{Kurkela:2014vha} three representative EoS parametrizations were obtained by an interpolating polytrope built from two “monotropes” of the form $p(n)=\kappa n^{\Gamma}$ matched in a smooth way, without considering any phase transition. 
The addition of a third polytropic segment (or more) would result in an increase of the region of allowed EoSs, although small in comparison with other uncertainties of the calculation. 
Taking into account a possible first-order phase transition would lead to an enlarged mass-radius region.
As a conservative approach we do not consider this possibility. 
We choose their EoSI and EoSII as they generate values of the tidal deformability close to the constraint coming from the GW170817 event \cite{Abbott_2018}. These EoSs are shown in Fig.~\ref{fig:EoSOM}. We moreover map these two EoSs to the EoSs for the inner crust \cite{NEGELE1973298} and the outer region \cite{PhysRevC.73.035804}, whereas for very low-densities ($\rho < 3.3 \cdot 10^3 {\rm g/cm^3}$) we use the Harrison-Wheeler EoS \cite{Harrison_Wheeler_Book}.

\begin{figure}[!h]
    \centering
    \includegraphics[scale=0.395]{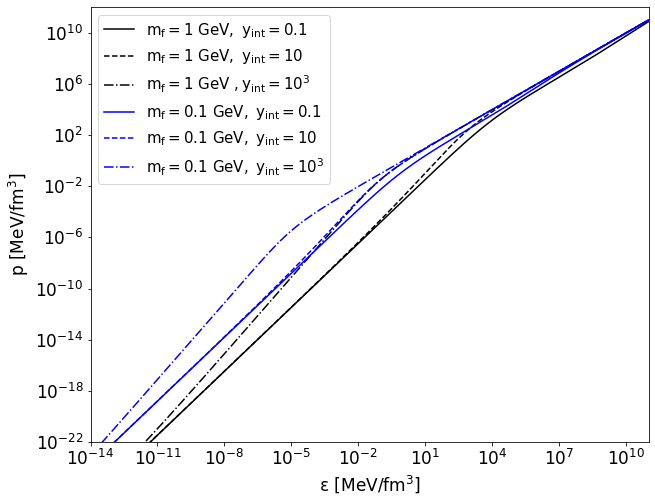}
    \caption{EoSs for fermionic self-interacting DM. Two different particle masses ($m_f=0.1$ GeV and $m_f=1$ GeV) and three different strength parameters ($y_{\rm int}=0.1,10,10^3$) are used.}
    \label{fig:EoSDM}
\end{figure}

 As for DM, we model it using one of the simplest model assumption, that is, as a non-selfannihilating self-interacting Fermi gas where the dimensionless energy density and pressure are given by \cite{Narain:2006kx}
\begin{eqnarray}
\epsilon_{DM}' &=&  \frac{1}{8 \pi ^2} \left[ \left(2 z^3 + z\right)\left(1+
  z^2\right)^{\frac{1}{2}} - \sinh ^{-1}\left(z\right) \right] \nonumber \\ &+&
\left(\frac{1}{3 \pi ^2}\right)^2 y^2 z^6  , 
\label{eqi:den1}\\
p_{DM}' &=& \frac{1}{24 \pi ^2} \left[ \left(2 z^3 - 3 z\right)\left(1+
    z^2\right)^{\frac{1}{2}} + 3  \sinh ^{-1}\left(z\right) \right] \nonumber \\ &+&
\left(\frac{1}{3 \pi ^2}\right)^2  
 y^2  z^6  
\label{eqi:pres1}
\end{eqnarray}
where $z$ is the dimensionless Fermi momentum and $y= m_{f}/m_{I}$ is defined as the ratio between the DM particle mass ($m_f$) and the interaction mass scale ($m_I$). Whereas for strong interactions
$m_I \sim 100$ MeV (the QCD scale, exchange of vector mesons), for weak interactions
$m_ I \sim 300$ GeV (electroweak scale, exchange of W and Z bosons). In Fig.~\ref{fig:EoSDM} we show six representative dimensionfull DM EoSs for DM particle masses of 0.1 GeV and 1 GeV, and three different strength parameters ($y_{\rm int}=0.1,1,10^3$). We will consider these different values throughout the text, hence moving from weakly to strongly self-interacting DM. 

In order to investigate compact objects made of OM admixed with DM, it is essential to carry out an analysis of the stable configurations. The computation of stable configurations for a single fluid can be found in Ref.~\cite{shapiro2008black}, where the stability for the different radial modes is analysed. For the study of a multifluid system we  follow the procedure of Ref.~\cite{Hippert:2022snq}, which we here briefly explain. 

First we need to compute the number of particles of each species, that in dimensionless units is obtained from 
\begin{equation}
    \dfrac{dN'_i}{dr}=4 \pi \dfrac{n'_i}{\sqrt{1-2(m_{OM}+m_{DM})/r}} r^2,
    \label{eq:Npart}
\end{equation}
where $n'_i=n_i/m_f^3$ with $i= \{ {\rm OM}, {\rm DM} \}$ is the dimensionless number density for each species and $N'_i$ is the dimensionless number of particles of the different fluids. We obtain the number of particles for each species by re-scaling them as $N_i=N_i' \cdot M_p^3/m_f^3$, with $M_p$ the Planck mass (see Ref.~\cite{Narain:2006kx}). These equations are coupled to the TOVs of Eqs.~(\ref{eq:TOV}) and have to be solved simultaneously.

As explained in Ref.~\cite{Hippert:2022snq}, one has to consider small radial perturbations of the equilibrium configuration by solving the Sturm-Liouville problem, whose solutions are determined by frequency eigenvalues. These eigenvalues follow a hierarchy. For a single fluid, when the first eigenvalue is $\omega_0^2<0$, the radial perturbation leads to an exponential growth, causing the instability. Therefore, the onset of instability corresponds to the point where 
$\omega_0^2$ becomes negative. At this point, small perturbations of the central energy density leave the number of particles unchanged. For a two-fluid configuration, the number of particles for OM and DM remains stationary under variations of each of the central densities $\epsilon^{ c}_{i}$, 
\begin{equation}
\small{
    \begin{pmatrix}
        \delta N_{OM} \\[0.75em]
        \delta N_{DM}
    \end{pmatrix} = 
    \begin{pmatrix}
        \partial N_{OM}/\partial \epsilon^{c}_{OM} & \partial N_{OM}/\partial \epsilon^{c}_{DM} \\[0.75em]
        \partial N_{DM}/\partial \epsilon^{c}_{OM} & \partial N_{DM}/\partial \epsilon^{c}_{DM}
    \end{pmatrix}
    \begin{pmatrix}
        \delta \epsilon^{c}_{OM} \\[0.75em]
        \delta \epsilon^{c}_{DM}
    \end{pmatrix} =0.}
    \label{eq:Matrix}
\end{equation}
For a single fluid, Eq.~(\ref{eq:Matrix}) leads to the solution $\partial N /\partial \epsilon^{c}=0$, which is equivalent to the criterion used for a single fluid, $\partial M/\partial p^{c}=0$~\cite{shapiro2008black,schaffner2020compact}.

For a two-fluid configuration, the matrix in Eq.~(\ref{eq:Matrix}) can be diagonalized obtaining two independent sets of variables, ($\epsilon^{c}_A,N_A$) and ($\epsilon^{c}_B,N_B$), with eigenvalues $\kappa_A$ and $\kappa_B$. Stable configurations can only happen when both eigenvalues, $\kappa_A$ and $\kappa_B$, are positive \cite{Hippert:2022snq}. This generalises the stability criterion for one fluid to two fluids. Note that in our previous works \cite{tolos2015dark,Dengler:2021qcq} a naive stability study was carried out based on a separate stability analysis of DM and OM. However, this analysis did not take into account the changes that one fluid could induce in the other, as discussed in Ref.~\cite{Hippert:2022snq}.

Also, we should comment that for a single fluid, the mass-radius relation is a curve and stable regions are separated by points, whereas for two fluids, the mass-radius relation can form  one or more areas, and the stable(s) region(s) are delimited by a set of limiting (or critical) lines . This is due to the fact that the matrix of Eq.~(\ref{eq:Matrix}) must have zero determinant so as non-trivial solutions exist, thus, leading to one condition for two independent variables $\epsilon^{c}_{OM}$ and $\epsilon^{c}_{DM}$. 

This analysis is equivalent to the one presented in Refs.~\cite{HENRIQUES1990511,PhysRevD.87.084040,BenKain2021, Diedrichs:2023trk}, where the stability curves are obtained by computing the contour lines for the total mass  and the number of particles in each of the two-fluid system. The limiting curves for the stable(s) region(s) are determined by finding the extrema of the total mass following contour lines for fixed particle numbers.

\subsection{Tidal Deformability}

The tidal deformability $\lambda$ is a physical quantity that  measures  the induced quadrupole moment, $Q_{ij}$, of a star due to the tidal field of the companion, $\mathcal{E}_{ij}$ \cite{Hinderer:2007mb,Hinderer_2010} as
\begin{equation}
Q_{ij}=-\lambda \mathcal{E}_{ij} .
\end{equation}
It is connected to the dimensionless second Love number $k_2$ \cite{Postnikov_2010,Hinderer_2010} as
\begin{equation}
    \lambda=\frac{2}{3}k_2 R^5 ,
\end{equation}
where $R$ is the radius of the star. The quantity $k_2$ can be calculated from
\begin{equation}
\begin{split}
    k_2=&\dfrac{8C^5}{5} (1-2C)^2 \left[ 2+2C(y_R-1)-y_R \right] \times \\
    & \{ 2C[6-3y_R+3C(5y_R-8)] + \\
    & 4C^3[13-11y_R+C(3y_R-2)+2C^2(1+y_R)]+ \\
    & 3(1-2C)^2[2-y_R+2C(y_R-1)]\ln (1-2C) \} ^{-1},
\end{split}
\label{eq:k2}
\end{equation}
where $C$ is the compactness. The compactness, after rescaling, is defined as $C=M_T/R$, with $M_T=M_{OM}+M_{DM}$ and $R=max(R_{OM},R_{DM})$.  We can calculate $y_R=y(R)$ by solving, together with the TOV of Eqs.~(\ref{eq:TOV}), the following equations
\begin{equation}
    r\dfrac{dy(r)}{dr}+y^2(r)+y(r) F(r) + r^2 Q(r)=0,
\label{eq:y}
\end{equation}
with
\begin{align}
    F(r)&=\dfrac{r-4\pi r^3 ((\epsilon'_{OM}+\epsilon'_{DM})-(p'_{OM}+p'_{DM}))}{r-2(m_{OM}+m_{DM})},
    \label{eq:F} \\
    Q(r)&=\dfrac{4 \pi r}{r-2(m_{OM}+m_{DM})} \nonumber \\
    &\times \left[ 5(\epsilon'_{OM}+\epsilon'_{DM}) +9(p'_{OM}+p'_{DM}) \right. \nonumber \\
    &+\left. \dfrac{\epsilon'_{OM}+p'_{OM}}{c^2_{s,OM}}+ \dfrac{\epsilon'_{DM}+p'_{DM}}{c^2_{s,DM}}- \dfrac{6}{4 \pi r^2} \right] \nonumber \\
    &-  4 \left[ \dfrac{(m_{OM}+m_{DM})+4\pi r^3 (p'_{OM}+p'_{DM})}{r^2 \left( 1-\dfrac{2(m_{OM}+m_{DM})}{r}\right)} \right] ^2,
    \label{eq:Q}
\end{align}
where $c_{s,i}(r)^2 = dp_i'/d\epsilon_i'$ is the squared speed of sound for  $i=$OM,DM and the initial condition for $y$ is given by $y(r=0)=2$ \cite{Postnikov_2010,Hinderer_2010}. With the second Love number we can finally compute the dimensionless tidal deformability by
\begin{equation}
    \Lambda = \dfrac{2 k_2}{3C^5}.
    \label{eq:TidalDef}
\end{equation}

\section{Results}
\label{sec:results}

\begin{figure*}[]
\begin{minipage}{0.45\linewidth}
    \centering
    \includegraphics[scale=0.4]{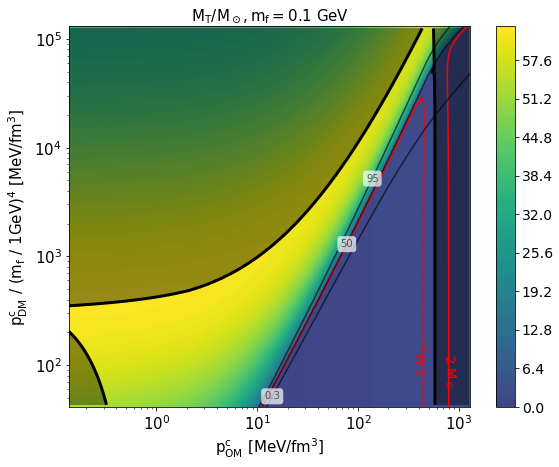}
\end{minipage}
\begin{minipage}{0.45\linewidth}
    \centering
    \includegraphics[scale=0.4]{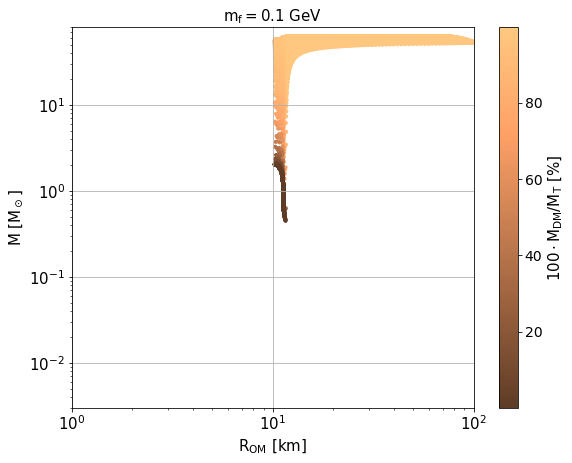}
\end{minipage}

\begin{minipage}{0.45\linewidth}
    \centering
    \includegraphics[scale=0.4]{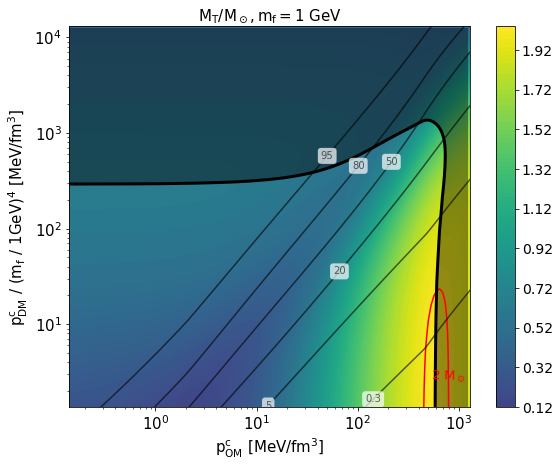}
\end{minipage}
\begin{minipage}{0.45\linewidth}
    \centering
    \includegraphics[scale=0.4]{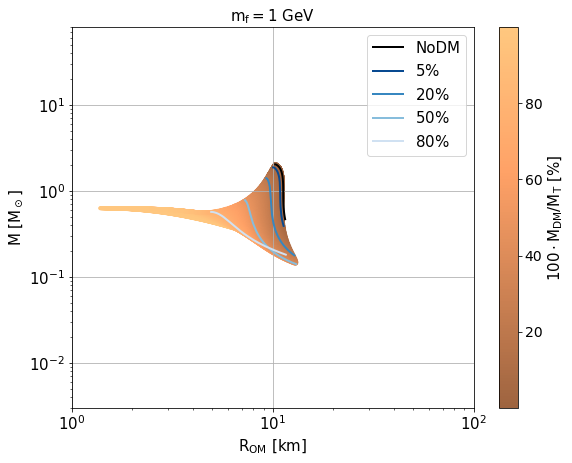}
\end{minipage}

\begin{minipage}{0.45\linewidth}
    \centering
    \includegraphics[scale=0.4]{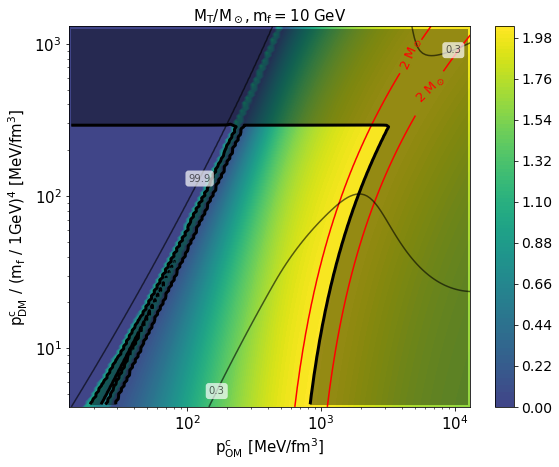}
\end{minipage}
\begin{minipage}{0.45\linewidth}
    \centering
    \includegraphics[scale=0.4]{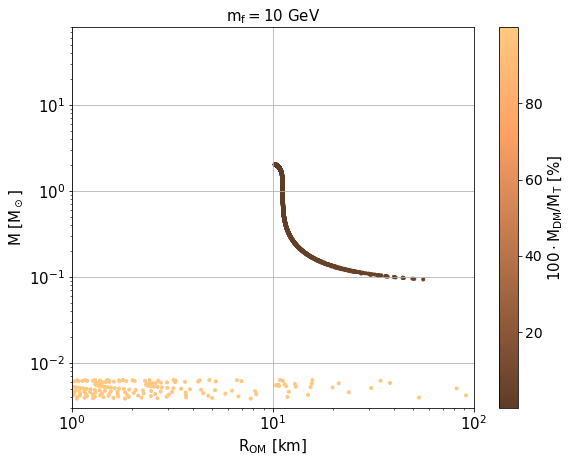}
\end{minipage}
\caption{Configurations of DM admixed with OM obtained for different DM particle masses at a fixed strength parameter of $y_{\rm int}=0.1$ and using EoSI. (Left plots) The total mass $M_T$ is shown as a function of the central pressures  $p^c_{OM}$ and $p^c_{DM}$ for different DM particle masses (different rows). The black lines represent the critical curves, and the unstable regions are shaded. The red line indicates the contour line for $M_{T}=2 M_\odot$. Some contour lines for different amounts of DM are also plotted in thin black solid lines. (Right plots) The mass-radius relation for the stable configurations (total mass as a function of OM radius) is shown. The different colored lines in the middle right plot indicate the mass-radius configurations with different fixed amount of DM ($100 \cdot M_{DM}/M_{\rm T}$). }
\label{fig:PressuresMR}
\end{figure*}

In this Section we present our results for the masses, radii and tidal deformabilities for compact stars that contain DM. We start by showing in Fig.~\ref{fig:PressuresMR} our stability analysis of the compact stellar objects admixed with DM and the mass-radius of the stable configurations. We fix the interaction strength parameter to $y_{\rm int}=0.1$ and take EoSI from Ref.~\cite{Kurkela:2014vha}, while considering different DM particle masses. We note that an analogous analysis can be performed for  other interaction strength parameters and the EoSII parametrization of Ref.~\cite{Kurkela:2014vha}. 

In the left plots of Fig.~\ref{fig:PressuresMR} we show contour plots for total mass $M_T$ as a function of the OM and DM central pressures for the different DM particle masses. Note that the central DM pressures are normalized to the DM particle mass. The shaded areas correspond to the unstable regions, according to the criterion discussed in Sec.~\ref{sec:formalism}. The red lines represent the contour lines for $M_{\rm T}=2M_\odot$, whereas the thin black solid lines depict the contour lines for different DM fractions. In the right plots of Fig.~\ref{fig:PressuresMR}, the mass-radius configurations for the different DM particle masses are plotted for only the stable cases, with the coloring indicating the amount of DM.

We start by analysing the top plots of Fig.~\ref{fig:PressuresMR}, where a low mass for the DM particle is considered ($m_f=0.1~{\rm GeV}$). In the top-left plot we observe a critical vertical line for $p_{OM}^c\approx 580~\rm{MeV}/\rm{fm}^3$. This  pressure corresponds to the critical pressure for a single fluid made of OM.  As discussed in Ref.~\cite{BenKain2021}, when one of the central pressures is much larger compared to the other one, the fluid with the largest central pressure dominates, and the configuration behaves as a single fluid. This is indeed our case here since the central pressures for DM are much smaller than the ones for OM in the region nearby this vertical line. The stable mass-radius configurations located to the left of this vertical line are plotted in the top-right plot of Fig.~\ref{fig:PressuresMR}, coloured according the DM content. As expected, the solutions close to the vertical line correspond to the solutions with masses up to 2$M_{\odot}$ and radii around 10 km, since these mass-radius configurations contain small DM fractions. When we move further left to this critical line in the ($p^c_{OM},p^c_{DM}$) plane, the DM content increases and new solutions appeared with larger masses and radii. As discussed in Ref.~\cite{Narain:2006kx}, the total mass of a single fermionic star increases as the particle mass decreases ($M\propto m_f^{-2}$). Therefore, for low DM particle masses and a significant DM content we can have configurations with masses up to $60~\rm{M}_\odot$, where a compact star made of OM is surrounded by a massive DM halo. We find that the larger the DM amount becomes, the larger the maximum masses are.

\begin{figure}[!h]
    \centering
    \includegraphics[scale=0.395]{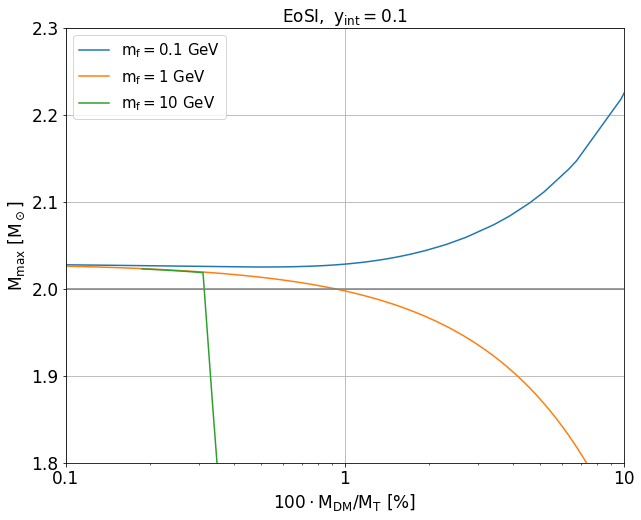}
    \caption{Maximum total mass ($M_{\rm max}$) for the stable configurations as a function of the amount of DM ($100 \cdot M_{\rm DM}/M_T$) for three DM particle masses ($m_f=0.1,1,10$ GeV). We consider the case of the DM interaction strength $y_{\rm int}=0.1$ and EoSI for OM. The grey line indicates the $2\rm{M}_\odot$ limit \cite{Demorest2010ShapiroStar,antoniadis2013massive,Fonseca2016,cromartie2020relativistic}.}
    \label{fig:MaxMass}
\end{figure}

As for the middle plots of Fig.~\ref{fig:PressuresMR}, we start with the analysis of the middle-left plot, where we display the stable and unstable configurations for $m_f=1{\rm GeV}$. As discussed previously, for low values of $p^c_{OM}$ or $p^c_{DM}$, the system behaves as one single fluid. In the middle-right plot, we show again the stable mass-radius solutions coloured according to the amount of DM, while depicting the mass-radius relations with solid blue lines for certain DM fractions. It is interesting to note that, for a mass of around 1 GeV, the increase of the DM fraction leads to a decrease in the maximum mass, as well as the reduction of the OM radii, in contrast to the case for masses below 1 GeV. This outcome was already found in Ref.~\cite{Deliyergiyev:2019vti} and will be discussed in more detail in Fig.~\ref{fig:MaxMass}, where we analyse the dependence of the total mass with the DM content.

In the bottom panels of Fig.~\ref{fig:PressuresMR}, the configurations with $m_f=10~{\rm GeV}$ are presented.  Note that two stable regions are seen below the critical horizontal line separated by diagonal lines in the ($p^c_{OM},p^c_{DM}$) plane. The one for low $p^c_{OM}$ corresponds to stable solutions where the OM pressures are close to those found in white dwarfs, whether the stable region at large $p^c_{OM}$ contains configurations close to the neutron star solutions.  The  horizontal line is located at $p_{DM}^c \approx 300 \cdot (m_f/1~{\rm GeV})^4~\rm{MeV}/\rm{fm}^3$. Hence, for the case of $m_f=10~{\rm GeV}$, matter behaves as a DM fluid in the region near this critical line. Indeed, the stable mass-radius configurations below but close to this horizontal line in the stable region for low $p^c_{OM}$  correspond to solutions for large DM fractions with masses around $10^{-2}~\rm{M}_\odot$, as seen in the bottom-right plot of Fig.~\ref{fig:PressuresMR}. The scattered points should be seen as an area of stable solutions. These are compatible with the dark compact planets described in Refs.~\cite{tolos2015dark,Deliyergiyev:2019vti,Dengler:2021qcq}. As we move away from the horizontal critical line to lower values of $p^c_{DM}$ while increasing $p^c_{OM}$, thus entering the stable region for large $p^c_{OM}$, the DM content is reduced and we recover the typical mass-radius configurations for neutron stars. We find that there is a  maximum for the amount of DM that can be accumulated in a compact star. The stable configurations with masses and radii similar to those of a neutron star contain very small DM cores, and increasing the amount of DM would make them unstable, as the DM core would collapse, as we will discuss in the next figure.

Once the stability of the mass-radius configurations with DM has been performed, we can now determine how the properties (mass, radius and tidal deformability) of the stable compact configurations change as we increase the amount of DM. First, we analyse how the maximum total mass ($M_{\rm max}$) for values close to the 2$M_{\odot}$ varies with increasing DM content ($100 \cdot M_{\rm DM}/M_T$), as shown in Fig.~\ref{fig:MaxMass} for the three different DM particle masses ($m_f=0.1,1,10$ GeV) and interaction strength $y_{\rm int}=0.1$. The calculations are performed with EoSI \cite{Kurkela:2014vha}. For a given DM particle mass, the values for $M_{\rm max}$ are obtained from the corresponding left panel in Fig.~\ref{fig:PressuresMR} by following the critical curves, as these limiting lines contain the information on the different extrema of the total mass, and in particular for values close to 2$M_{\odot}$. The DM fraction for a given $M_{\rm max}$ is determined by the crossing between the critical curve and a given contour line for the DM content.

As seen in Fig.~\ref{fig:MaxMass}, there are three different behaviours for $M_{\rm max}$ as a function of the amount of DM, depending on the DM particle mass. For low masses ($m_f= 0.1~\rm{GeV}$), the maximum mass increases with the amount of DM. In this scenario, DM accumulates in a halo around the neutron star, without compromising its stability, leading to the formation of massive configurations, as already discussed in Fig.~\ref{fig:PressuresMR}. For intermediate masses ($m_f=1~\rm{GeV}$), the maximum masses decrease. In these configurations, a DM core grows, while the OM radius and mass decrease, as seen in the right-middle plot of Fig.~\ref{fig:PressuresMR}. For high masses ($m_f=10~\rm{GeV}$), the behaviour is notably different. Similar to  the intermediate case, a DM core is formed, but when it reaches a certain mass, it becomes unstable, producing the instability of the whole configuration, as mentioned in the discussion of Fig.~\ref{fig:PressuresMR}. The DM fraction at which this fast change occurs depends on the DM particle mass and the interaction strength. Note that the maximum mass of a pure DM compact star, i.e. the onset of stability for a single-fluid object, is proportional to $y/m_f^2$~\cite{Narain:2006kx} \footnote{Note that for some configurations where the DM fraction is very large and, hence, the star becomes purely a DM compact object, the maximum mass is higher than $2\rm{M}_\odot$, fulfilling again the constraints coming from observations.} While this analysis can be repeated using other strength parameters and other EoSs for OM, the behaviour of the maximum masses is similar for the different cases, only depending on the  DM particle mass. The maximum mass tends to increase at small values of $m_f$, and the abrupt change of stability of the DM core would still occur at high values of the DM particle mass.

\begin{figure}[!h]
    \centering
    \includegraphics[scale=0.38]{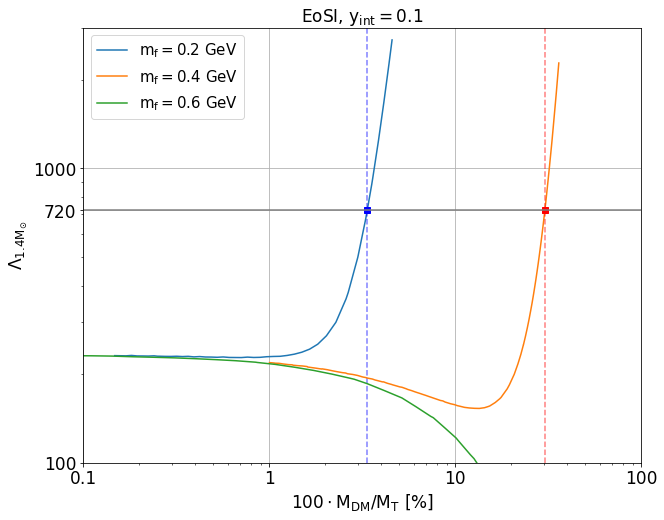}
    \caption{ Tidal deformability of $M=1.4 M_\odot$  star ($\Lambda_{1.4 M_{\odot}}$)  star as a function of the amount of DM ($100 \cdot M_{\rm DM}/M_T$) for three DM particle masses ($m_f=0.2, 0.4, 0.6$ GeV). We consider the case of the DM interaction strength $y_{\rm int}=0.1$ and EoSI for OM. The grey line indicates the upper limit for the tidal deformability obtained from GW170817 ($\Lambda_{1.4}<720$)~\cite{Abbott_2018}. The vertical dashed lines show the amount of DM that is needed to obtain solutions outside $\Lambda_{1.4}\le 720$~\cite{Abbott_2018} for each configuration.}
    \label{fig:TidalEoSI}
\end{figure}

\begin{figure}
    \centering
    \includegraphics[scale=0.38]{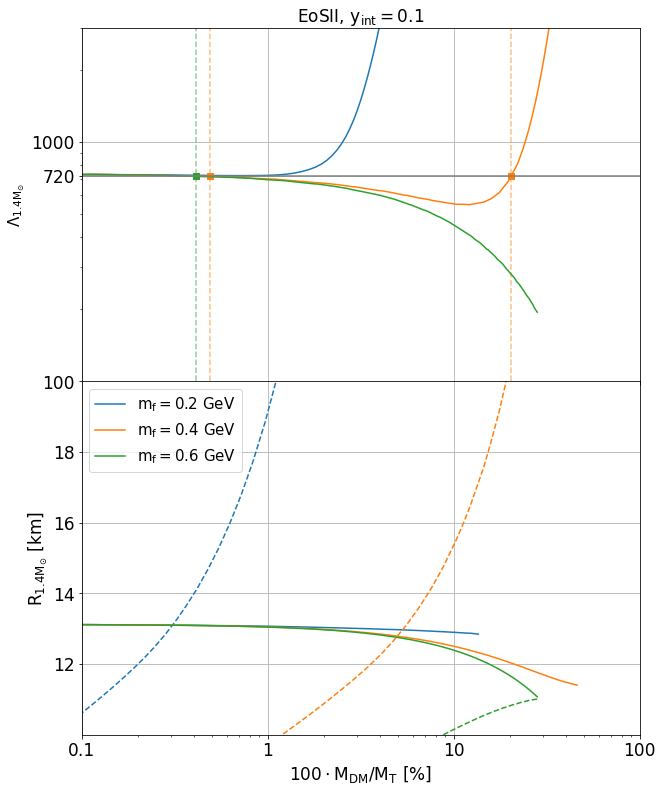}
    \caption{In the top panel the tidal deformability of $M=1.4 M_\odot$ star ($\Lambda_{1.4 M_{\odot}}$) is shown whereas the corresponding radii ($R_{1.4 M_{\odot}}$) for OM (solid lines) and DM (dashed lines) are displayed in the low panel, both quantities as function of the amount of DM ($100 \cdot M_{DM}/M_T$) for three DM particle masses ($m_f=0.2,0.4,0.6$ GeV). We consider the case of the DM interaction strength $y_{\rm int}=0.1$ and EoSII for OM. Again, the vertical dashed lines show the amount of DM that is needed to obtain solutions outside $\Lambda_{1.4}\le 720$ \cite{Abbott_2018} for each configuration. 
    }
    \label{fig:TidalEoSII}
\end{figure}

 Apart from the changes in the total mass, the radius and, hence, the tidal deformability will also vary depending on the DM fraction. As mentioned earlier, for low masses of DM particles, a halo is created around neutron stars. As the radius of this halo grows, the tidal deformability changes. For the analysis we have followed contour lines of $M=1.4M_\odot$ in the ($p^c_{OM},p^c_{DM}$) plane, and plotted in Fig.~\ref{fig:TidalEoSI} the tidal deformability of these objects versus the DM fraction, obtained from the crossings between the $M=1.4M_\odot$ contour line and those for different DM content. In this figure we consider EoSI for OM and the strength parameter for DM  $y_{\rm int}=0.1$. The behaviour is similar for the smallest DM particle masses ($m_f=0.2,0.4)$ GeV, that is,  the DM radius increases until it reaches the OM radius, creating a halo and consequently increasing the tidal deformability~\cite{Nelson_2019}. As for $m_f\gtrsim 0.6$ GeV,  the tidal deformability stays  below $\Lambda_{1.4}=720$~\cite{Abbott_2018}.

In fact, the behaviour of the tidal deformability with DM masses using OM EoSI is similar to the one for OM EoSII. In the upper panel of Fig.~\ref{fig:TidalEoSII} we show the tidal deformability for a $M=1.4M_\odot$ star whereas in the low panel we display the corresponding radii ($R_{1.4 M_{\odot}}$) for OM (solid lines) and DM (dashed lines), both quantities as functions of the DM content. In a configuration without DM, EoSII provides a tidal deformability of $\Lambda_{1.4}\approx 738$, that is  slightly  higher than the limit derived from GW170817 of $\Lambda_{1.4}<720$~\cite{Abbott_2018}. However, increasing the fraction of DM can lead to results where $\Lambda_{1.4}\lesssim 720$, as seen in Ref.~\cite{Ellis:2018bkr}. This outcome can be understood by analysing the behaviour of the radii for OM and DM. In the lower panel, for small masses of DM particle ($m_f=0.2~$GeV), the DM halo is created very quickly, increasing the tidal deformability. For larger masses ($m_f=0.4~$GeV and $m_f=0.6~$GeV), as we increase the DM fraction while maintaining the total mass constant to 1.4$M_{\odot}$, the OM mass decreases. This decrease in OM mass leads to a reduction in the OM radius, whereas the DM radius increases steadily. Thus, the tidal deformability for 1.4$M_{\odot}$ stays below 720 for a larger content of DM as compared to the $m_f=0.2~$GeV case, agreeing with the GW170817 observation.

Up to now we have discussed the behaviour of the maximum mass and the tidal deformability as we increase the DM fraction for different DM particle masses and a fixed value of the DM interaction strength. We have payed a special attention on how the presence of DM might induce deviations of these properties from the 2$M_{\odot}$ observations  \cite{Demorest2010ShapiroStar,antoniadis2013massive,Fonseca2016,cromartie2020relativistic} and tidal deformability extracted from GW170817 event~\cite{Abbott_2018}. We have considered two EoSs that give rise to a large set of solutions compatible with these two previous observations. Then, to conclude this Section, we perform an exhaustive analysis for the maximum mass and tidal deformability, by means of considering the two EoS parametrizations and not only taking into account the different DM particle masses but also varying the interaction strength $y_{\rm int}$. In this manner we are able to constrain the amount of DM that can be accumulated in compact stars while still fulfilling the aforementioned observations.

\begin{figure*}[]

    \centering
    \includegraphics[scale=0.55]{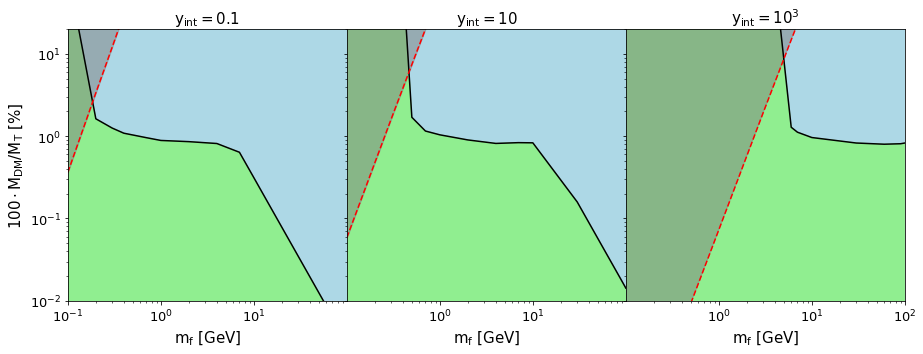}
\caption{Amount of DM ($100\cdot M_{DM}/M_T$) as a function of its particle mass ($m_f$) for different interaction strength parameters $y_{\rm int}$ (different columns for $y_{\rm int}=0.1,10,10^3$) using EoSI for OM. The black curves show the DM fractions that produce maximum stable masses with $M=2\ \rm{M}_\odot$ ~\cite{Demorest2010ShapiroStar,antoniadis2013massive,Fonseca2016,cromartie2020relativistic}.  The regions below these limiting curves indicate solutions with masses above $M=2\ \rm{M}_\odot$. The red dashed lines give the DM fraction with $\Lambda_{1.4}=720$. The area to the right of the dashed red lines are stable configurations with tidal deformabilities below 720~\cite{Abbott_2018}.}
\label{fig:EosI}
\end{figure*}

\begin{figure*}[]

    \centering
    \includegraphics[scale=0.55]{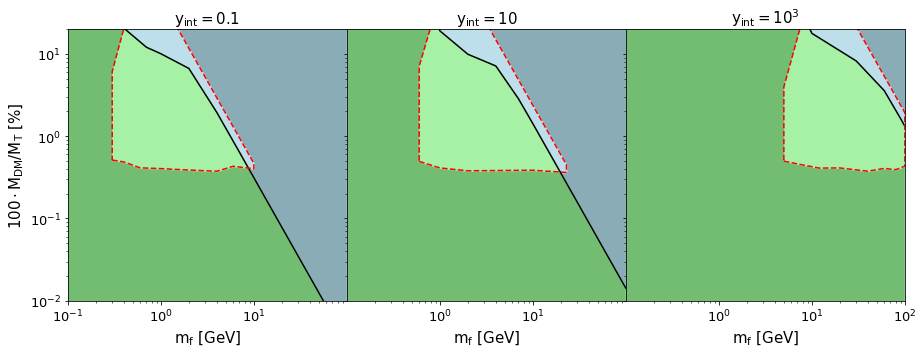}
\caption{The same as Fig.~\ref{fig:EosI} but for OM EosII. Whereas the regions below the black solid curves indicate solutions with $M \geq 2\ \rm{M}_\odot$~\cite{Demorest2010ShapiroStar,antoniadis2013massive,Fonseca2016,cromartie2020relativistic},  the red dashed lines encapsulate the areas where $\Lambda_{1.4}\le 720$~\cite{Abbott_2018}. The region being compatible with the astrophysical constraints extends to the unphysically limiting case of pure DM compact stars, as those configurations could reach 2$M_{\odot}$ and give rise to small tidal deformabilities.
}

\label{fig:EosII}
\end{figure*}

In Figs.~\ref{fig:EosI},\ref{fig:EosII} we show the amount of DM ($100\cdot M_{DM}/M_T$) as a function of the DM particle mass ($m_f$) for the strength parameter $y_{\rm int}=0.1$ (left panels), $y_{\rm int}=10$ (middle panels) and $y_{\rm int}=10^3$ (right panels). Results for EoSI for OM are shown in Fig.~\ref{fig:EosI}, whereas the outcome for EoSII
is displayed in Fig.~\ref{fig:EosII}. 
The black solid curves indicate the fraction of DM with a maximum stable mass of $M=2M_\odot$. The regions below these lines are composed of stable configurations with $M \geq 2 M_\odot$. The dashed red lines, on the other hand, are stable configurations with tidal deformabilities equal to 720. Whereas in Fig.~\ref{fig:EosI} the area to the right of these red lines conforms stable configurations with tidal deformabilities below 720, in Fig.~\ref{fig:EosII} the red dashed lines encapsulate the areas where $\Lambda_{1.4}\le 720$.

In Fig.~\ref{fig:EosI} we find a different behaviour of the maximum mass and tidal deformability depending on the DM particle mass, as already discussed in the Figs.~\ref{fig:MaxMass}, \ref{fig:TidalEoSI} for $y_{\rm int}=0.1$. For large masses, we have a DM core that becomes unstable. As we decrease the DM particle mass, the region of maximum masses above 2$M_{\odot}$, that the core can reach, becomes larger as the amount of DM that can be admixed in these configurations increases. For low masses, the amount of DM tends to 100\% due to the creation of a DM halo around the compact star. The constraint coming from GW170817 only affects low masses, for which the DM halo is created as the maximum radius increases.

Fig.~\ref{fig:EosII} is similar to Fig.~\ref{fig:EosI} but obtained using EoSII. The most notorious difference is the quantity of DM allowed by the tidal deformability constraint, as the stable configurations with tidal deformabilities below 720 are encapsulated by the red dashed lines. This outcome can be understood by looking at Fig.~\ref{fig:TidalEoSII} and the left panel in Fig.~\ref{fig:EosII}, both obtained for $y_{\rm int}=0.1$. In this case, for small DM mass particles, a large DM halo is created and, hence, the tidal deformability has values above 720 for all DM fractions.  For intermediate masses between approximately 0.3 GeV and 10 GeV, the tidal deformability reaches values below 720 for DM fractions above 5\%  as the OM radius decreases whereas the DM halo has not been formed yet. For larger masses, DM becomes unstable and no solutions are found. Similar allowed DM contents are obtained for larger interaction strengths but the encapsulated region is shifted to larger DM particle mass with increasing $y_{\rm int}$.

Moreover, we can extract some conclusions by comparing the DM fractions as a function of the DM particle mass for the two EoSs. For large DM particle masses, the instability region is independent of the OM EoS used. The slope of the 2$M_{\odot}$ line in the high DM mass region in Figs.~\ref{fig:EosI},\ref{fig:EosII} is caused by the DM core that becomes unstable, whose mass only depends on the DM particle mass and interaction strength. As mentioned earlier, in these cases, since $p_{DM}^c$ is much higher that $p_{OM}^c$, the stability of the DM core dominates the stability of the whole configuration.  Regarding low DM particle masses, the limiting 2$M_{\odot}$ line is shifted to larger DM fractions when using EoSII. As this EoS generates higher masses, the DM fraction needed to reduce the maximum mass to 2$M_{\odot}$ would be also be larger. 

 As mentioned in the Introduction, our results should be compared to the outcome of Refs.~\cite{Ivanytskyi:2019wxd,Mariani:2023wtv,Karkevandi:2024vov,liu2024dark}. We should first note that our present work is focused on determining possible stable solutions close and beyond the neutron star configurations, while Refs.~\cite{Ivanytskyi:2019wxd,Mariani:2023wtv} former studies concentrate on masses and radii close to the ones of neutron stars. With regard to these latter mass-radius solutions, our work and the ones of Refs.~\cite{Ivanytskyi:2019wxd,Karkevandi:2024vov,liu2024dark} aim at determining the amount of either fermionic or bosonic DM that can be accumulated in neutron stars taking into account certain astrophysical constraints, such as  2$M_{\odot}$ observations and the tidal constraint coming from the GW170817 event. In the case of Ref.~\cite{Ivanytskyi:2019wxd} a rough estimate of the accreted DM into neutron stars in the most central region of the Galaxy is used and in Ref.~\cite{liu2024dark} the DM model is constrained from observational limits imposed on the DM self-interaction cross section. A similar trend for the DM fraction with DM particle masses and interaction strengths is observed in all works. However, the calculations in Refs.~\cite{Ivanytskyi:2019wxd,Karkevandi:2024vov,liu2024dark} have been based on a specific model for the OM EoS, so the conclusions are model dependent, as compared to our analysis based on two representative OM EoSs which delineate the lower and upper limits of the mass-radius curves of OM. Moreover, there is no mention on the stability of the solutions, as we have discussed at length in our present work.  Concerning Ref.~\cite{Mariani:2023wtv}, the authors have chosen two EoSs, one \textit{soft} and one \textit{stiff}, to serve as the limiting cases, representing an envelope for numerous microscopic hadronic EoSs. They have constrained DM as a self-interacting Fermi gas, depending on the DM particle mass, interaction strength parameter $y_{int}$ and a general coupling for the self-interaction $g$ (for self-annihilation), based on observations for the mass, radius and tidal deformability of neutron stars as well as the accepted cosmological dark matter freeze-out values and self-interaction cross-sections obtained from galactic dynamics. Whereas this former study is also aiming at providing model-independent results for the DM content as ours, the discussion on the stability analysis of the different mass-radius configurations is missing, which is particularly important for large DM particle masses when the DM core becomes unstable.

\subsection{Massive Compact Stars}

Up to now we have discussed the amount of DM that can be accumulated in compact stars while still fulfilling the $2M_{\odot}$ and tidal deformability constraints considering the two limiting OM EoSs. However, from Fig.~\ref{fig:PressuresMR} it is clear that some stable solutions with DM reach $60M_{\odot}$ for the case of EoSI. Therefore, it is interesting to explore whether more massive configurations than $2M_{\odot}$ can be obtained while still fulfilling the tidal deformability constraint, in particular in view of two recent observations, that is, the black widow pulsar PSR J0952-0607 of $2.35 \pm 0.17 M_{\odot}$  \cite{Romani:2022jhd} and the compact object with mass $2.5-4.5 M_{\odot}$ detected in the gravitational wave event GW230529$\_$18150  \cite{LIGOScientific:2024elc}.

\begin{figure}[!h]
    \centering
    \includegraphics[scale=0.395]{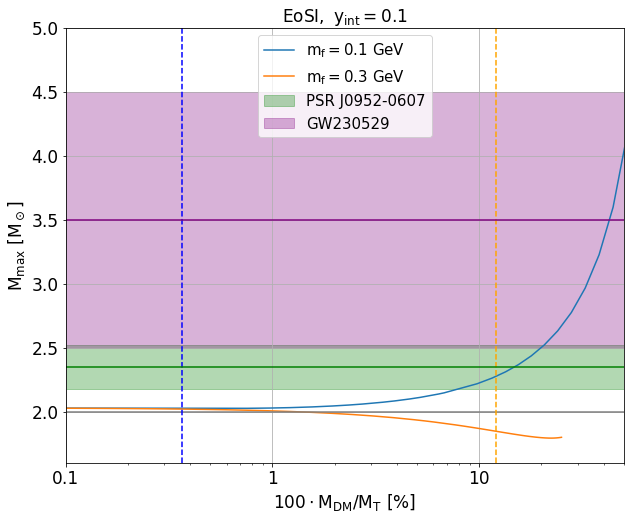}
    \caption{Maximum mass as a function of the amount of DM, using EoSI for OM and $y_{\rm int}=0.1$ for DM, and two DM particle masses (0.1 GeV and 0.3 GeV). The dashed vertical lines indicate the maximum amount of DM compatible with $\Lambda_{1.4}<720$ for each particle mass. The green and purple areas represent the masses in PSR J0952-0607 and GW230529$\_$18150, respectively.}
    \label{fig:EoSI_GW230529}
\end{figure}

 In order to see whether we can find stable solutions with DM content with masses similar to PSR J0952-0607 or GW230529$\_$18150 while still fulfilling the tidal deformability constraint, in Fig. \ref{fig:EoSI_GW230529} we show the maximum mass as a function of the amount of DM for two DM particle masses (0.1 GeV and 0.3 GeV) using EoSI. The dashed vertical lines indicate the maximum amount of DM compatible with $\Lambda_{1.4}<720$ for each particle mass. As shown in this figure, it is not possible to find solutions with masses close to either the pulsar PSR J0952-0607 or the heaviest compact object in the GW230529$\_$18150 merger while still fulfilling the tidal deformability constraint. For $m_f=0.1$ GeV the maximum amount of DM allowed by the tidal constraint is compatible with masses $\sim 2 M_{\odot}$, whereas for $m_f=0.3$ GeV the DM content can go above 10$\%$ but the maximum mass of the compact object remains below $2 M_{\odot}$. For smaller or larger particle masses as well as larger interaction strengths, the situation is similar 
 so that no solutions with masses above $2M_{\odot}$ compatible with the tidal constraint are found.
 
\begin{figure}[!h]
    \centering
    \includegraphics[scale=0.395]{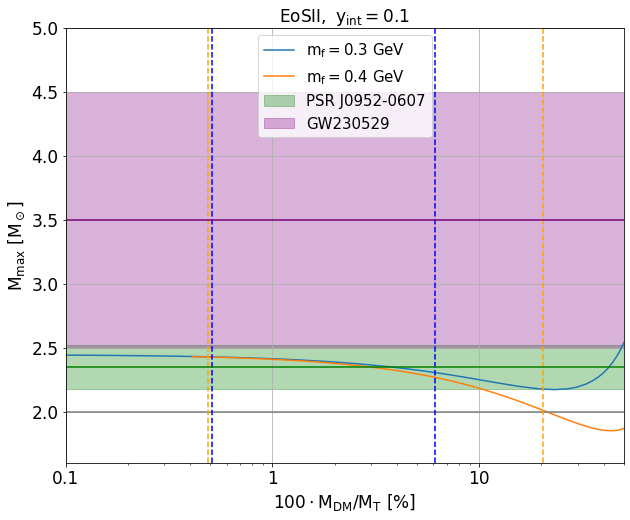}
    \caption{Maximum mass as a function of the amount of DM, using EoSII for OM and $y_{\rm int}=0.1$ for DM, and two DM particle masses (0.3 GeV and 0.4 GeV). The dashed vertical lines indicate the minimum and maximum amount of DM compatible with $\Lambda_{1.4}<720$ for each particle mass. The green and purple areas represent the masses in PSR J0952-0607 and GW230529$\_$18150 respectively.}
    \label{fig:EoSII_GW230529}
\end{figure}

 With regards to the case of considering EoSII, in Fig.~\ref{fig:EoSII_GW230529} we observe that for $m_f=$0.3 GeV or 0.4 GeV the mass of PSR J0952-0607 can be reached while the tidal constraint is still fulfilled, as the DM amount associated with this maximum mass falls between the minimum and maximum limits in DM content determined by the tidal constraint, indicated by the dashed vertical lines for each mass. However, maximum masses larger than $2.5 M_{\odot}$ cannot be obtained, thus no solutions with masses in the $2.5-4.5 M_{\odot}$ mass range of the compact object in GW230529$\_$18150 are found. For larger or smaller masses and larger interaction strengths,  no solutions with masses close to the  PSR J0952-0607 mass compatible with the tidal constraint are obtained.

\section{Summary}
\label{sec:summary}

In this paper we have performed a comprehensive study of compact stars admixed with non-self annihilating self-interacting DM. We have started from two parametrized EoSs for OM that fulfil the well-known limits at low and large nuclear densities while giving rise to masses and radii that span the possible mass-radius region compatible with 2$M_\odot$ observations and the constraint on the tidal deformability coming from the GW170817 event, hence being representative of
a wide variety of EoS that fall between these two limiting cases. We aim at determining the amount of DM that can be accumulated in compact stars while still fulfilling these previous astrophysical constraints by means of a controlled scheme for OM, thus going beyond the use of individual models.

To this end, we have first carried out the analysis of the possible mass-radius stable configurations following the procedure of Ref.~\cite{Hippert:2022snq}. In that work the authors take into account the fact that any variation in the stability of one of the systems would affect the other, and vice versa. Thus, we have improved on our previous naive stability analysis of two separate OM and DM fluids of Refs.~\cite{tolos2015dark,Dengler:2021qcq}. In this manner, we have obtained stable mass-radius configurations of OM and DM with radii smaller than 10 km and masses similar to Earth-like or Jupiter-like stellar objects, as already reported in Refs.~\cite{tolos2015dark,Dengler:2021qcq}, whereas new configurations have appeared for very low DM particle mass, such as those configurations consisting of OM surrounded by a massive DM halo, with masses reaching up to $60~\rm{M}_\odot$.

Then, we have studied how maximum masses around 2$M_\odot$ and tidal deformabilities for $1.4M_{\odot}$ stars vary as a function of the DM fraction for different DM particle masses and interaction strengths considering the two EoS parametrizations.  Whereas there is a strong dependence on the maximum masses close to 2M$_\odot$ with the DM particle mass, the tidal deformability for $1.4M_{\odot}$ depends on both the DM particle mass and the EoS. As a general trend, we find that for large DM particle masses above few tenths of GeV, no stable solutions compatible with the astrophysical observables are found as the DM core becomes unstable. As the DM particle mass is reduced, more and more stable configurations appear compatible with 2$M_{\odot}$ as the DM fraction increases. The specific values of the DM fraction depend on the self-interaction strength and the EoS.  For the soft nuclear EoSI, the DM fraction is limited to 10\%. Typically, for masses ranging from 0.1 to 10 GeV, the DM fraction can be at most 1\%. For a stiff nuclear EoS, such as EoSII, the DM fraction can be increased to 10\% or more. However, the mass is constrained to be between 0.3 GeV and 10 GeV for the weak self-interacting case and has to be at least 5 GeV for the strong self-interacting DM. For DM particle masses of less than 0.1 GeV, stable neutron star configurations with more than 1\% of self-interacting DM are ruled out by the constraint of the tidal deformability from GW170817 irrespective on the chosen  limiting nuclear EoS.

 To finalize our summary, some remarks are in order on the recent observations of compact stars with more than $2M_{\odot}$, such as the black widow pulsar PSR J0952-0607 and the compact object in the mass gap detected in the GW230529$\_$18150 event. Although large mass compact stars with DM can be found, our study indicates that it is difficult to reconcile masses above $2.5 M_{\odot}$ with the tidal deformability constraint from GW170817, even when including a possible fermionic self-interacting DM component in the neutron star.

\begin{acknowledgments}
We thank the anonymous referee for instructive comments and raising the issue about the maximum possible mass.
M.F. Barbat acknowledges support from CSIC under JAE Intro 2023 program (No.~JAEINT$\_$23$\_$EX$\_$0032). This work was also supported under contract No.~PID2022-139427NB-I00 financed by the Spanish MCIN/AEI/10.13039/501100011033/FEDER,UE and from the project CEX2020-001058-M Unidad de Excelencia ``Mar\'{\i}a de Maeztu''. Moreover, this project has received funding from the European Union Horizon 2020 research and innovation programme under the program H2020-INFRAIA-2018-1, grant agreement No.\,824093 of the STRONG-2020 project, from the CRC-TR 211 'Strong-interaction matter under extreme conditions'- project Nr. 315477589 - TRR 211 and from the Generalitat de Catalunya under contract 2021SGR00171.
\end{acknowledgments}

\newpage
\bibliography{refs}
\bibliographystyle{apsrev4-1}

\end{document}